\documentclass[hyper, letterpaper]{JHEP3}
\usepackage{amsmath, amsthm}
\usepackage{amssymb, amscd}
\usepackage{mathrsfs}
\usepackage[all]{xy}
\usepackage{amsbsy}
\usepackage{bm}
\usepackage{graphicx}
\usepackage{epstopdf}
\usepackage{verbatim}
\usepackage{multirow}
\usepackage{longtable}

\numberwithin{equation}{section}

\newcommand{\eg}{\emph{e.g.}}
\newcommand{\ie}{\emph{i.e.}}
\newcommand{\cf}{\emph{cf.}}
\newcommand{\ol}{\overline}

\newcommand{\be}{\begin{equation}}
\newcommand{\ee}{\end{equation}}
\newcommand{\ben}{\begin{equation*}}
\newcommand{\een}{\end{equation*}}
\newcommand{\bea}{\begin{eqnarray}}
\newcommand{\eea}{\end{eqnarray}}
\newcommand{\bean}{\begin{eqnarray*}}
\newcommand{\eean}{\end{eqnarray*}}
\newcommand{\nno}{\nonumber}

\newcommand{\bTi}{\begin{itemize} \setlength{\itemsep}{-.1cm}}
\newcommand{\eTi}{\end{itemize}}

\newcommand{\pd}{\partial}

\newcommand{\Tr}{{\rm Tr}}

\newcommand{\Vol}{{\rm Vol}}

\newcommand{\bA}{{\bf A}}
\newcommand{\bE}{{\bf E}}
\newcommand{\bU}{{\bf U}}

\newcommand{\ev}[1]{\langle {#1} \rangle}

\newcommand{\R}{{\mathbb{R}}}
\newcommand{\C}{{\mathbb{C}}}
\newcommand{\Z}{{\mathbb{Z}}}
\newcommand{\F}{{\mathbb{F}}}
\newcommand{\Q}{{\mathbb{Q}}}
\renewcommand{\L}{{\mathbb{L}}}

\renewcommand{\P}{{\mathbb{P}}}

\newcommand{\Hom}{{\rm Hom}}

\newcommand{\Ext}{{\rm Ext}}

\newcommand{\CC}{{\cal C}}

\newcommand{\CE}{{\cal E}}

\newcommand{\CH}{{\cal H}}

\newcommand{\CK}{{\cal K}}

\newcommand{\CN}{{\cal N}}
\newcommand{\CO}{{\cal O}}

\newcommand{\CR}{{\cal R}}

\newcommand{\CW}{{\cal W}}

\newcommand{\CZ}{{\cal Z}}

\newcommand\bes{\begin{equation*}}
\newcommand\ees{\end{equation*}}
\newcommand{\beas}{\begin{eqnarray*}}
\newcommand{\eeas}{\end{eqnarray*}}

\newcommand{\bse}{\begin{subequations}}
\newcommand{\ese}{\end{subequations}}

\newcommand{\rf}{{\rm ref}}
\newcommand{\mot}{{\rm mot}}
\newcommand{\e}[1]{{\hat{e}_{#1}}}

\newtheorem{rmk}{Remark}
\newtheorem{prp}{Proposition}

\title{Quantum Wall Crossing in $\CN=2$ Gauge Theories}

\author{Tudor Dimofte,$^1$ Sergei Gukov,$^{1,2}$ and Yan Soibelman$^{3}$
\\
\\
$^1$ California Institute of Technology, Pasadena, CA 91125, USA \\
$^2$ University of California, Santa Barbara, CA 93106, USA\\
$^3$ Department of Mathematics, KSU, Manhattan, KS 66506, USA}

\abstract{We study refined and motivic wall-crossing formulas
in $\CN=2$ supersymmetric gauge theories with $SU(2)$ gauge group
and $N_f < 4$ matter hypermultiplets in the fundamental representation. Such gauge theories provide an excellent testing ground for the conjecture that ``refined $=$ motivic.''
}

\preprint{CALT 68-2766}

\begin{document}


\section{Introduction}
\label{sec:intro}

Much can be said about the quantum physics of a supersymmetric system
by looking at the spectrum of its BPS states.
In the present paper, we take a closer look at the spectrum of BPS states
in $\CN=2$ supersymmetric gauge theories in four dimensions.
These theories serve as an excellent laboratory for testing
various predictions for wall-crossing behavior of
the refined BPS invariants
recently proposed in \cite{DG}, which carry information not only about the charge of BPS states but about their spin content.
In particular, we are able to test the general proposal that refined = motivic.

As a byproduct of our study of wall-crossing formulas
in $\CN=2$ supersymmetric gauge theories, we discover
new mathematical identities for the quantum dilogarithm function
\begin{equation}
\label{qdilog}
\bE(x) := \sum_{n=0}^\infty \frac{(-q^{\frac12}x)^n}{(1-q)\cdots(1-q^n)} = \prod_{i=0}^\infty\big(1+q^{i+\frac12}x\big)^{-1}\,.
\end{equation}
The quantum dilogarithm function ${\bf E} (x)$ has many remarkable properties.\footnote{See section 3.3 of \cite{DGLZ}
for a recent review in the closely related context of quantization
of the moduli space ${\cal M}_{{\rm flat}} (G_{\C}, \Sigma)$
of flat $G_{\C}$ connections on a Riemann surface $\Sigma$.}
Perhaps one of the most beautiful and well-known properties is the so-called ``pentagon'' identity
\begin{equation}
\label{eeeee}
{\bf E} (x_1) {\bf E} (x_2) = {\bf E} (x_2) {\bf E} (x_{12}) {\bf E} (x_1)\,,
\end{equation}
where $x_1 x_2 = q x_2 x_1$ and $x_{12} = q^{-1/2} x_1 x_2 = q^{1/2} x_2 x_1$.
This identity describes a basic wall-crossing process,
in which two hypermultiplet states
with primitive charge vectors $\gamma_1$ and $\gamma_2$, and
with symplectic product $\langle \gamma_1, \gamma_2 \rangle = 1$,
form a bound state of total charge $\gamma = \gamma_1 + \gamma_2$.
On one side of the wall of marginal stability, corresponding
to the right-hand side of \eqref{eeeee}, the bound state is stable.
On the other side of the wall, the bound state decays and
the space of single-particle BPS states
contains only the two stable particles of charge $\gamma_1$ and $\gamma_2$,
represented by the factors ${\bf E} (x_1)$ and ${\bf E} (x_2)$
on the left-hand side of \eqref{eeeee}. \\

\begin{figure}[h]
\centering
\includegraphics[width=4in]{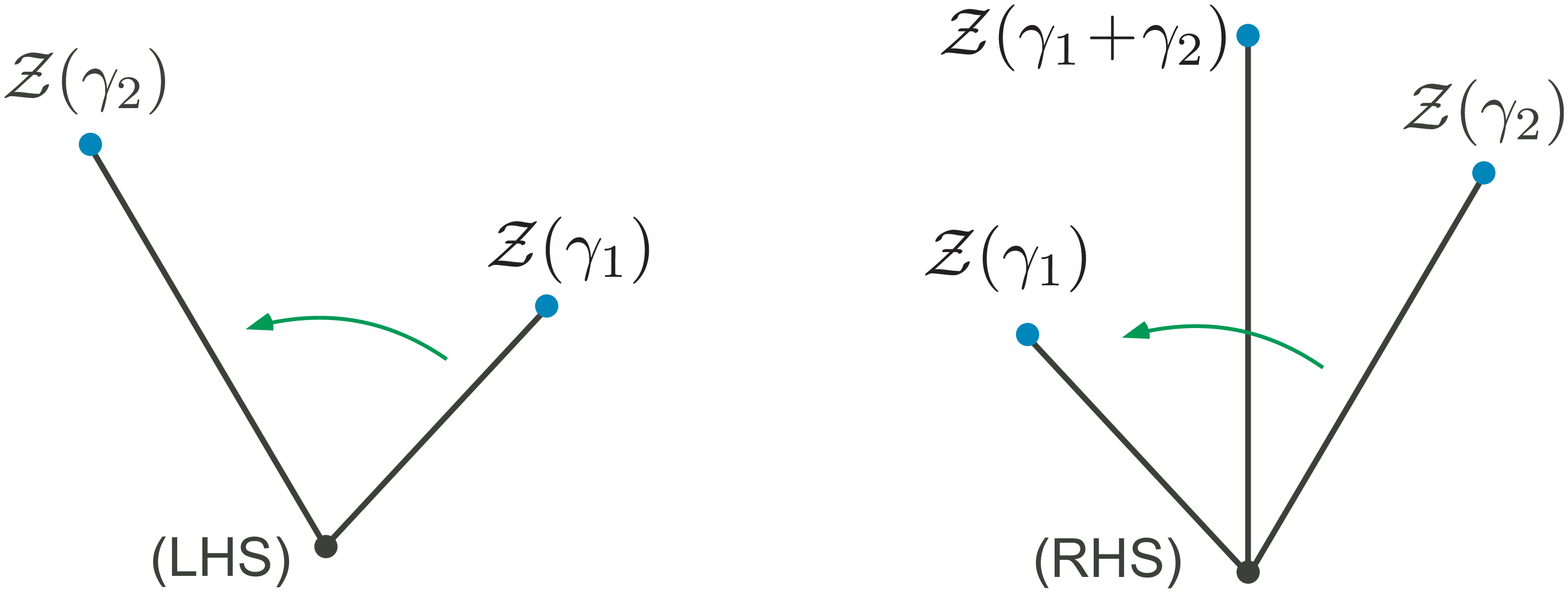}
\caption{Visualization of the wall crossing encoded in the pentagon identity,
in terms of BPS rays in a central charge plane.
The presence of the bound state $\gamma_1+\gamma_2$ depends on
the relative arguments of the central charges $\CZ(\gamma_1)$ and $\CZ(\gamma_2)$.}
\label{fig:K1}
\end{figure}

The pentagon identity \eqref{eeeee} is the first
relation in an entire tower of operator identities obeyed
by the quantum dilogarithm function,
which correspond to $\langle \gamma_1, \gamma_2 \rangle = k$
with arbitrary values of $k \ge 1$.
For example, in the next simplest case $k=2$, the following identity holds:
\begin{equation} \label{puresym}
N_f=0:\quad \bU_{2,-1}\bU_{0,1} = \bU_{0,1}\bU_{2,1}\bU_{4,1}\cdots\bU^{\rm vect}_{2,0}\cdots\bU_{6,-1}\bU_{4,-1}\bU_{2,-1}\,,
\end{equation}
where we used the shorthand notation $\bU_{n,m} : = {\bf E} (q^{-\frac{nm}{2}} x_1^n x_2^m)$
and
\begin{equation} \label{Uvect}
\bU^{\rm vect}_{2,0} := \bE(-q^{\frac12} x_1^2)^{-1}\bE(-q^{-\frac12} x_1^2)^{-1}\,.
\end{equation}
The identity \eqref{puresym} encodes the wall crossing
in pure $\CN=2$ super-Yang-Mills theory with gauge group $SU(2)$.
Notice that its left-hand side is similar to that of \eqref{eeeee},
whereas the right-hand side is now an infinite product.
Just like in the pentagon identity, the two sides of the equality \eqref{puresym}
correspond to two chambers separated by a wall of marginal stability,
and each factor $\bU_{n,m}$ represents a stable BPS state of electric charge $n$ and magnetic charge $m$
($\bU^{\rm vect}_{2,0}$ corresponds to a vector multiplet of electric charge $2$).

Besides the basic identity \eqref{puresym} we find the following identities
\begin{align}
N_f = 1: \quad& \bU_{1,-1}\,\bU_{1,0}\,\bU_{0,1} = \bU_{0,1}\,\bU_{1,1}\,\bU_{2,1}\,\bU_{3,1}\cdots\bU_{1,0}^2\,\bU_{2,0}^{\rm vect}\cdots \bU_{3,-1}\,\bU_{2,-1}\,\bU_{1,-1} \nonumber \\
N_f = 2: \quad& \bU_{1,-1}^2\,\bU_{0,1}^2 = \bU_{0,1}^2\,\bU_{1,1}^2\,\bU_{2,1}^2\,\bU_{3,1}^2\cdots\bU_{1,0}^4\,\bU_{2,0}^{\rm vect}\cdots \bU_{3,-1}^2\,\bU_{2,-1}^2\,\bU_{1,-1}^2\, \label{qSW} \\
N_f =3:\quad& \bU_{1,{-2}}\,\bU_{0,1}^4 = \bU_{0,1}^4\,\bU_{1,2}\,\bU_{1,1}^4\,\bU_{3,2}\,\bU_{2,1}^4\cdots\bU_{1,0}^6\,\bU_{2,0}^{\rm vect}\cdots\bU_{2,-1}^4\,\bU_{3,-2}\,\bU_{1,-1}^4\,\bU_{1,{-2}}\,, \nonumber
\end{align}
which describe the spectrum of BPS states, with spin content,
in the $SU(2)$ Seiberg-Witten theory with $N_f = 1,2,3$ fundamental matter multiplets.
Note that the classical limit, $q \to 1$, of equations \eqref{puresym} and \eqref{qSW}
describes the BPS spectrum without spin content; for the $\CN=2$ theories in question,
it has been discussed in both the mathematics \cite{KS1} and physics \cite{GMN,GMN-II} literature.
More generally, every $\CN=2$ gauge theory leads to identities
like \eqref{puresym} - \eqref{qSW}, one for each wall of marginal stability.
It would be interesting to extend the analysis of the present
paper to study refined wall crossing
in $\CN=2$ gauge theories with more general gauge groups and matter content.

We begin this paper with an example-based discussion of BPS spectra and wall crossing in $\CN=2$ gauge theories with gauge group $SU(2)$. The identities \eqref{puresym}-\eqref{qSW} are first motivated directly via gauge theory, and then interpreted geometrically, in terms of BPS states on a Calabi-Yau three-fold. In Sections \ref{sec:refined}-\ref{sec:math}, we then re-evaluate formulas \eqref{puresym}-\eqref{qSW} in the general context of refined, quantum, and motivic wall crossing. In particular, we show how these formulas may be derived via the motivic Donaldson-Thomas invariants of \cite{KS1}.


\section{$\CH_{BPS}$ in $SU(2)$ gauge theory with $N_f$ flavors}
\label{sec:hbps}

We wish to study the space of BPS states, $\CH_{BPS}$,
in $\CN=2$ supersymmetric gauge theories in four dimensions.
Specifically, we revisit pure $\CN=2$ super-Yang-Mills with gauge group $SU(2)$ and $\CN=2$ supersymmetric QCD with $N_f \leq 3$ matter hypermultiplets in the fundamental representation.
We will focus on the case where these hypermultiplets have vanishing bare masses prior to symmetry breaking.

The BPS spectra of $\CN=2$ gauge theories with and without matter were first considered in the seminal papers of Seiberg and Witten \cite{SWi, SWii}. Brane constructions \cite{WittenM} and geometric engineering \cite{Lerche-noncrit, KKV} of $\CN=2$ gauge theories provided a further interpretation of BPS states as geodesics (or ``curves of constant central charge'') on the Seiberg-Witten curve. Here, we begin by reviewing BPS spectra and wall crossing purely in the gauge theory context. We then show that gauge theory spectra are also related to collections of stable vector bundles on appropriate complex surfaces. We motivate this correspondence via geometric engineering, which supplies the dictionary between BPS states in gauge theory and BPS invariants of Calabi-Yau three-folds (which contain these complex surfaces).

\subsection{Coulomb branches and BPS states}
\label{sec:coul}

Let us recall some basic properties of
$SU(2)$, $\CN=2$ gauge theories with $N_f\leq 3$ flavors of massless fundamental matter. At low energies, the gauge group of these theories is broken to an abelian subgroup $U(1)\subset SU(2)$.
The corresponding Coulomb branches are then parametrized by a single complex modulus $u$,
the expectation value of the Casimir of the the complex adjoint scalar in the $\CN=2$ vector multiplet: $u = \ev{\Tr\,\phi^2}$.
We draw the basic structure of the Coulomb branches for $N_f=0,1,2,3$ further below in Figure \ref{fig:Coul}.
Since the coefficient of the one-loop beta-function is $N_f-2N_c=N_f-4$,
the supersymmetric gauge theory is asymptotically free when $N_f<4$.
With the expectation value $|u|$ thought of as the energy scale of the theory,
the large-$|u|$ region of the Coulomb branch corresponds to weak coupling,
while the small-$|u|$ region corresponds to strong coupling.
These two regions are separated by a wall of marginal stability.

The quantum properties of these theories are encoded in the geometry of their Seiberg-Witten curves.
For $N_f$ flavors, we have \cite{SWi, SWii} (see also \cite{HOz})
\be \Sigma_{SW}^{(N_f)}\,\,:\qquad y^2 = (x^2-u)^2-\Lambda_{N_f}^{4-N_f}x^{N_f} \,,\label{Sigma_SW} \ee
where $\Lambda_{N_f}$ is the strong-coupling scale, determined by the one-loop beta function. (The monomial $x^{N_f}$ would be deformed to a product $\prod_i(x-m_i)$ if bare masses were turned on.) The Seiberg-Witten differential on this curve is a meromorphic one-form satisfying
\be \pd_u\lambda_{SW} \sim \frac{dx}{y}\,. \notag \ee
Note that for vanishing bare masses, the genus-one curve $\Sigma_{SW}$ may have ``punctures'' --- in particular, there is a puncture for every fundamental hypermultiplet --- but the Seiberg-Witten differential has no residues at these punctures \cite{SWii}.

Since the low-energy theory is abelian with rank-one gauge group $U(1)$,
the charge lattice $\Gamma = \Z\oplus\Z$ is two-dimensional.
It is generated by an elementary electric charge $\gamma_e = (1,0)$ and an elementary magnetic charge $\gamma_m = (0,1)$.
This charge lattice is identified with the homology of the Seiberg-Witten curve with all punctures filled in,
\be \Gamma \simeq H_1(\ol{\Sigma}_{SW},\Z)\,. \ee
The central charge of any state with a charge $\gamma\in\Gamma$ is then given by the integral of $\lambda_{SW}$ along the corresponding one-cycle $\gamma$ on the Seiberg-Witten curve:
\be \CZ(\gamma) = \oint_\gamma\lambda_{SW}\,. \label{Z_gamma} \ee
In particular, letting $\gamma_e$ and $\gamma_m$ be a canonical basis of cycles on $\Sigma_{SW}$ and setting
\be a(u) = \oint_{\gamma_e} \lambda_{SW}\,,\qquad a_D(u) = \oint_{\gamma_m}\lambda_{SW}\,, \ee
the central charge of a state with electric charge $n$ and magnetic charge $m$ is
\be \CZ(n,m;u) = a(u)n+a_D(u)m\,.\ee
Recall that (by definition) the mass of a BPS state is determined by the absolute value of its central charge:
\be M(\gamma) \sim |\CZ(\gamma)|\,. \label{mbps} \ee

Note that, in these $\CN=2$ theories, the massive $W^\pm$~bosons
have electric charge that is twice that of the fundamental electric charge
(\ie\ twice the charge of a fundamental electric hypermultiplet).
It is the fundamental charge that we call $\gamma_e$. At large $|u|$,
the central charges $a(u)$ and $a_D(u)$ are then determined by the one-loop beta function to be of the form
\begin{align} a(u) &= \frac12\sqrt{2u}+\ldots \\
 a_D(u) &= i\frac{4-N_f}{4\pi}\sqrt{2u}\log\frac{u}{\Lambda^2_{N_f}} + \ldots\,,
\end{align}
where the subleading terms are non-perturbative instanton corrections.

The structure of the Coulomb branches for the asymptotically-free $SU(2)$ theories
are depicted schematically in Figure \ref{fig:Coul}.
In each case, the weak-coupling and strong-coupling regions are separated by a single wall of marginal stability $\CW$.
This wall is defined by the condition
\be \arg a(u) = \arg a_D(u)\,,\qquad \text{or}\qquad \frac{a(u)}{a_D(u)}\in\R^+\,. \label{SWMS} \ee
Due to the BPS mass formula \eqref{mbps}, BPS states are allowed to combine or decay into other BPS states
when the arguments of the central charges of electric and magnetic states align as in \eqref{SWMS}.

\begin{figure}[htb]
\centering
\includegraphics[width=4.25in]{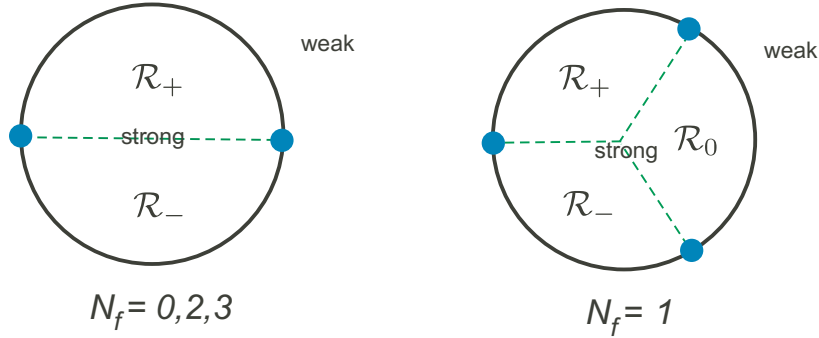}
\caption{The approximate structure of the Coulomb branch ($u$-plane) of $SU(2)$ Seiberg-Witten theories (\cf\ \cite{BF}). A wall of marginal stability separates strong and weak coupling. The blue dots correspond to singularities where BPS states become massless.}
\label{fig:Coul}
\end{figure}

Each Coulomb branch in Figure \ref{fig:Coul} has either two or three singular points
where dyonic hypermultiplet BPS states become massless.
These are located at values of $u$ on the discriminant locus $\Delta_{SW}$
where the Seiberg-Witten curve becomes singular.
It turns out that in the strong-coupling region the two or three BPS states
corresponding to these singular points are the \emph{only} stable BPS states.
Due to the non-local nature of the charge lattice $\Gamma$
(it is really a local system, fibered over the Coulomb branch),
there is no unique description of the charges of these BPS states.
In some of the different regions $\CR_+,\CR_-,\CR_0$ indicated in Figure \ref{fig:Coul},
the strong-coupling spectra can be described as (see \cite{BF}):
\bse \label{specS}
\begin{align} N_f = 0: \quad& [\CR_+]\;\; (2,1)\,,\;(0,1)\qquad [\CR_-]\;\; (-2,1)\,,\;(0,1)  \\
N_f=1: \quad& [\CR_0]\;\; (0,1)\,,\;(-1,1)\,,\;(1,0)\qquad  [\CR_+]\;\;(2,-1)\,,\;(-1,1)\,,\;(1,0) \nno \\ &\qquad  [\CR_-]\;\;(0,1)\,,\;(1,1)\,,\;(1,0) \\
N_f=2: \quad& [\CR_+]\;\; (1,-1)_2\,,\; (0,1)_2\qquad   [\CR_-]\;\; (1,1)_2\,,\; (0,1)_2   \\
N_f=3: \quad& [\CR_+]\;\; (-1,2)\,,\; (1,-1)_4\qquad   [\CR_-]\;\; (-1,2)\,,\; (0,1)_4\,.
\end{align}
\ese
The notation list states in terms of their electric and magnetic charges $(q,p)$;
subscripts denote multiplicities. Moreover, we have chosen a particle-antiparticle splitting,
and listed only the ``particles'' (the full spectrum must be completed by adding antiparticles).

At weak coupling (large $|u|$), the spectrum must be invariant under the monodromy of the charge lattice around $u=\infty$,
\be M_\infty\,:\,\,(n,m) \mapsto (n+N_f\,m,m)\,. \ee
The spectrum ends up being generated by the action of $M_\infty$ on the hypermultiplet states
corresponding to singularities, with the addition of the vector multiplets containing the massive $W^\pm$~bosons.
(Note that the action of $M_\infty$ on the electric states such as the $W^\pm$ is trivial.)
Altogether, one finds weak-coupling spectra
\bse \label{specW}
\begin{align} N_f = 0:\quad & (2,0)\,,\; (2n,1)\,,\quad n\in \Z\,,  \\
 N_f = 1:\quad& (2,0)\,,\; (1,0)_2\,,\; (n,1)\,,\quad n\in\Z\,, \\
 N_f = 2:\quad& (2,0)\,,\; (1,0)_4\,,\; (n,1)_2\,,\quad n\in\Z\,, \\
 N_f = 3:\quad& (2,0)\,,\; (1,0)_6\,,\; (n,1)_4\,\;(2n+1,2)\,,\quad n\in\Z\,.
\end{align}
\ese

Now the reader should be able to recognize the wall-crossing formulas
in pure $\CN=2$ super-Yang-Mills \eqref{puresym} and in $\CN=2$ SQCD \eqref{qSW}
stated in the introduction:
\begin{align*}
N_f=0:\quad& \bU_{2,-1}\bU_{0,1} = \bU_{0,1}\bU_{2,1}\bU_{4,1}\cdots\bU^{\rm vect}_{2,0}\cdots\bU_{6,-1}\bU_{4,-1}\bU_{2,-1} \\
N_f = 1: \quad& \bU_{1,-1}\,\bU_{1,0}\,\bU_{0,1} = \bU_{0,1}\,\bU_{1,1}\,\bU_{2,1}\,\bU_{3,1}\cdots\bU_{1,0}^2\,\bU_{2,0}^{\rm vect}\cdots \bU_{3,-1}\,\bU_{2,-1}\,\bU_{1,-1} \\
N_f = 2: \quad& \bU_{1,-1}^2\,\bU_{0,1}^2 = \bU_{0,1}^2\,\bU_{1,1}^2\,\bU_{2,1}^2\,\bU_{3,1}^2\cdots\bU_{1,0}^4\,\bU_{2,0}^{\rm vect}\cdots \bU_{3,-1}^2\,\bU_{2,-1}^2\,\bU_{1,-1}^2 \\
N_f =3\,:\quad& \bU_{1,{-2}}\,\bU_{0,1}^4 = \bU_{0,1}^4\,\bU_{1,2}\,\bU_{1,1}^4\,\bU_{3,2}\,\bU_{2,1}^4\cdots\bU_{1,0}^6\,\bU_{2,0}^{\rm vect}\cdots\bU_{2,-1}^4\,\bU_{3,-2}\,\bU_{1,-1}^4\,\bU_{1,{-2}}
\end{align*}
In each case, the strong-coupling spectrum appears on the left and the weak-coupling spectrum appears on the right. Or, to be more precise, exactly \emph{half} of the spectra appear: in order to write down these relations, we had to impose a split between particles and antiparticles. On each side of these formulas, the operators $\bU_{m,n}$ corresponding to the BPS particles (\emph{vs.} antiparticles) in the spectrum are listed in increasing order of the argument of their central charges. When the wall of marginal stability at $\arg a(u) = \arg a_D(u)$ is crossed, this ordering is reversed --- relating the two sides.

Note that the choice of splitting between particles and anti-particles that was made here is not particularly relevant. One can divide the charge lattice $\Gamma$ in half in many different ways, resulting in formulas that are simply related via conjugation by some (typically) finite set of operators $\bU_{m,n}$. In terms of the central charge plane, a change of splitting rotates a distinguished sector of total angle $\pi$, removing BPS states from one side of this sector and adding them to the other --- hence conjugation by the corresponding operators in the wall-crossing formula. For example, in the case $N_f=1$, the LHS of formula \eqref{qSW} corresponds to region $\CR_0$ at strong coupling. By changing the particle-antiparticle split we can obtain equivalent formulas that correspond to regions $\CR_+$ and $\CR_-$:
\begin{align*}
N_f=1\;[\CR_+]:\quad& \bU_{2,-1}\,\bU_{1,0}\,\bU_{-1,1} = \bU_{-1,1}\,\bU_{0,1}\,\bU_{1,1}\,\bU_{2,1}\,\bU_{3,1}\cdots\bU_{1,0}^2\,\bU_{2,0}^{\rm vect}\cdots \bU_{3,-1}\,\bU_{2,-1} \\
N_f=1\;[\CR_-]:\quad& \bU_{0,-1}\,\bU_{1,0}\,\bU_{1,1} = \bU_{1,1}\,\bU_{2,1}\,\bU_{3,1}\cdots\bU_{1,0}^2\,\bU_{2,0}^{\rm vect}\cdots \bU_{3,-1}\,\bU_{2,-1}\,\bU_{1,-1}\,\bU_{0,-1}\,.
\end{align*}
(A series of pentagon identities is used to rewrite conjugates of the LHS of \eqref{qSW} in the forms appearing here.)

The validity of the $N_f=0$ formula \eqref{puresym} can be verified in several different ways.
In Section \ref{sec:math}, we shall explain how it results from the study of motivic
Donaldson-Thomas invariants and quiver representations. The remaining formulas for $N_f>0$ can actually be derived from the $N_f=0$ case
by simply applying the pentagon identity \eqref{eeeee} repeatedly in the form
\be \bU_{\gamma_1} \bU_{\gamma_2} = \bU_{\gamma_2} \bU_{\gamma_1 + \gamma_2} \bU_{\gamma_1}
\qquad \text{for $\ev{\gamma_1 ,\gamma_2}=1$}\,. \ee
Thus, in terms of quantum wall crossing, one might say that the BPS spectrum
of $N_f=0$ $SU(2)$ theory ``predicts'' the spectra for $N_f>0$!

\subsection{Stable bundles on $\F_0$}
\label{sec:stab}

It is curious to note that the weak-coupling spectrum~(\ref{specW}a)
of stable BPS states in the pure $\CN=2$ super-Yang-Mills theory is closely related
to the collection of stable vector bundles --- or, to be more precise, coherent sheaves ---
on the Hirzebruch%
\footnote{We use the subscripts $b$ and $f$ on the factors here to denote ``base'' and ``fiber,'' in line with a more general geometric description coming in Section \ref{sec:string}.} %
surface $\F_0 = \P^1_b \times \P^1_f$\,.
Recall that a stable holomorphic vector bundle (or sheaf) $\CE$
on a complex surface $S$ is distinguished by the condition
\be \mu(\CE') < \mu(\CE) \label{slopestab} \ee
for all nontrivial proper sub-bundles $\CE'\subset\CE$.
Here, $\mu (\CE)$ is the slope of $\CE$, defined as
\be \mu(\CE) = \frac{1}{r} \int_S J \wedge c_1(\CE) \,, \ee
where $r$ is the rank of $\CE$ and $J$ is the K\"ahler form of $S$.

Topologically, a coherent sheaf $\CE$ is described by the Mukai vector
\be \gamma = {\rm ch} (\CE) \sqrt{\widehat{A} (S)} \,, \ee
which takes values in $H^{{\rm even}} (S; \Z)$.
In the present case of $S = \F_0$, this is a rank-4 lattice.
Anticipating the connection with $SU(2)$ gauge theory (\cf\ \eqref{Dem} in Section \ref{sec:string}),
we wish to restrict to a rank-2 sublattice $\Gamma \subset H^{{\rm even}} (S; \Z)$
characterized by the conditions
\be \int_{\P^1_f} c_1(\CE) = 0\,,\qquad \int_{\F_0} {\rm ch}_2(\CE)=0\,. \ee
In other words, the topological type of such a restricted bundle (or sheaf) $\CE$
is described by its rank $r$ (which we identify with magnetic charge $m$)
and
\be n = \int_{\P^1_b} c_1(\CE) \,, \ee
(which we identify with \emph{twice} the electric charge in $SU(2)$ gauge theory).
Note that such bundles with $\gamma = (n,m) \in \Gamma$ have zero discriminant
\be \Delta(\CE) = \frac{1}{2r^2}(c_1^2-2r\,{\rm ch}_2) =0 \,,\ee
so that the formula for the expected dimension of their moduli space
takes a very simple form (\cf\ \cite{DFR-noncpt, Diac-M} or \cite{LePotier}):
\be d = 1+2r^2\Delta-r^2\chi(\CO_{\F_0})=1-r^2\,. \ee
This dimension must be non-negative for any stable bundles to exist.
Therefore, there are two basic choices to consider, corrsponding to $r=1$ and $r=0$.
In the first case, all line bundles with $\gamma (\CE) \in \Gamma$ are stable, and
they correspond to the spectrum of dyons in $SU(2)$ gauge theory at weak coupling.
(The vanishing dimension $d=0$ implies that the classical moduli spaces of these states are points,
making them hypermultiplets.)

Similarly, in the case $r=0$ it is easy to see that the only stable sheaf
on $\F_0$ with $\gamma (\CE) \in \Gamma$ is a torsion sheaf, corresponding to the trivial rank-1 bundle $\CE = \CO (0)$ on $\P^1_b$.
Its classical moduli space is $\P^1_f$
(in agreement with $d=1$ from the dimension formula), which implies that the corresponding BPS states
form a vector multiplet. Of course, this is just the electric $W$~boson.
Altogether, we find the following spectrum of stable bundles (sheaves)
on the Hirzebruch surface $\F_0$ with Mukai vector $\gamma (\CE) \in \Gamma$\,:
\be \gamma = (1,0)\,,\; (n,1)\,,\quad n\in \Z\,, \ee
This is the complete set of charges~(\ref{specW}a) of stable BPS states in the weak-coupling spectrum
of the pure $\CN=2$ super-Yang-Mills theory.
A similar conclusion holds for other $\CN=2$ gauge theories with $N_f = 1,2,3$
fundamental matter multiplets, whose BPS spectra (\ref{specW}b)-(\ref{specW}d) correspond to the collection of stable holomorphic vector bundles (sheaves) on a complex surface $S$
obtained by blowing up $\F_0$ at $N_f$ generic points.
As we explain next, this is not an accident.

\subsection{Realization via BPS states on a Calabi-Yau 3-fold}
\label{sec:string}

In \cite{DG}, we described wall crossing for BPS states in $\CN=2$ compactifications of type II string theory on a Calabi-Yau three-fold $X$. It is well known that the $\CN=2$ gauge theories considered in this paper have such a realization in string theory, via geometric engineering \cite{KKV}. The stable BPS states in gauge theory then correspond to stable bound states of BPS D-branes on the Calabi-Yau $X$.

This correspondence leads to interesting alternative interpretations of BPS spectra and wall crossing in gauge theory. In particular, it is well known that the finite strong-coupling spectrum of $SU(2)$ theory is determined by the conifold singularities in the vector-multiplet moduli space of $X$. We will also see, as per the previous section, that the infinite weak-coupling spectrum is determined by the classically stable vector bundles on $X$.
The interpretation of BPS states as stable D-branes on a Calabi-Yau --- and especially as stable holomorphic branes in type IIA string theory --- also provides an important starting point for the connection between refined and motivic invariants considered in later sections.

An $\CN=2$ $SU(N)$ gauge theory with matter can be geometrically engineered in type IIA string theory by ``compactifying'' on a noncompact Calabi-Yau three-fold $X$ that is the total space of the anti-canonical line bundle over a complex surface $S$ \cite{KKV, KMV-SW},
\be X = \CO(-K)\to S\,. \nno\ee
For example, in the case of pure $SU(2)$ gauge theory, this surface can be taken to be the Hirzebruch surface $\F_0$\,:
\be SU(2),\,\,N_f=0:\qquad S = \F_0\,. \nno\ee
More generally, $SU(2)$ theory with $N_f\leq 3$ flavors of fundamental matter is constructed by blowing up $N_f$ generic points on $\F_0$.

Although $\F_0$ is just a simple product, $\F_0=\P^1\times\P^1$, it is convenient to treat it as a trivial fibration and to distinguish the two $\P^1$ factors as fiber and base. The area of (say) the ``base'' $\P^1$, in string units, is then proportional to the bare gauge theory coupling defined at the string scale, ${\rm Vol}(\P^1_b) \sim \frac{1}{g^2}$, while the area of the ``fiber'' $\P^1$ is proportional to the mass of the $W$ bosons, ${\rm Vol}(\P^1_f) \sim m_{W}\ell_s$. In order to send both the string scale and the four-dimensional Planck scales to infinity (to decouple stringy and gravitational physics and retain a pure gauge theory with finite $m_W$ and renormalized coupling), one must therefore consider an infinitely large base and an infinitesimally small fiber. Specifically, one can take base and fiber K\"ahler classes
\be
\begin{array}{rl}
\exp(2\pi i t_b) &\sim \Lambda^4\epsilon\,, \\ 2\pi it_f &\sim -\epsilon a\,, \end{array}
\qquad \epsilon \to 0\,.
\ee
(As usual, we define $t = B +i\Vol$.) The Coulomb branch of the resulting gauge theory, parametrized by $u$, is identified with a complex codimension-one slice of the K\"ahler moduli space of $X$ in the neighborhood of the ``Seiberg-Witten point''  $(e^{2\pi i t_b}, e^{2\pi it_f})\sim(0,1)$.

BPS states in gauge theory correspond to type IIA D-branes wrapping even cycles in $X$ (and extending along the time direction in $\R^4$). In the string realization, the full lattice of D-brane charges is identified\footnote{One should be careful considering $H^*(X)$ for a noncompact Calabi-Yau manifold. Being more precise, we could (\eg) take a compactification $\hat{X}$ of $X$ and consider the part of $H^*(\hat{X})$ that is dual to cycles in $H_*(\hat{X})$ that come from $H_*(X)$.} with $H^{\rm even}(X;\Z)\simeq \Z^4$. The lattice of gauge theory charges is a sublattice of $H^{\rm even}(X;\Z)$,
\be \Gamma \subset H^{\rm even}(X;\Z)\,, \ee
and includes only those states that remain light in the geometric engineering limit.

For pure $SU(2)$ theory with $N_f=0$ engineered on $\F_0\simeq \P^1_b\times \P^1_f$, it is clear that the $W$ boson comes from a D2-brane wrapped on a fiber. It turns out (see the end of this section) that the dual fundamental magnetic charge corresponds to a D4-brane wrapped on the entire surface $\F_0$:
\be \begin{array}{rl} 
 2\gamma_e &\,\,\leftrightarrow\,\, [\P^1_f]\,, \\
 \gamma_m &\,\,\leftrightarrow\,\, [\F_0]\,. \end{array} \label{Dem}
\ee
The electromagnetic product of the two charges is consistent with the intersection product in (co)homology,
\be \ev{2\gamma_e,\gamma_m} = -\int_X[\P^1_f]\wedge[\F_0] = -\int_{\F_0}[\P^1_f]\wedge[K_{\F_0}] = 2\,. \ee
Note that on $X = \CO(-K)\to\F_0$ there are \emph{no} compact cycles with intersection number $\pm 1$: all intersections are multiples of $2$. Likewise, in pure $SU(2)$ gauge theory with $N_f=0$, all electric charges are multiples of $\gamma_{W^\pm}=\pm2\gamma_e$.

Equivalently, one can engineer gauge theory in type IIB string theory by compactifying on the mirror Calabi-Yau $\tilde{X}$. Local mirror symmetry \cite{KKV, CKYZ} determines $\tilde{X}$ to have the form
\be \zeta\tilde{\zeta} = H(z,w) \ee
for $\zeta,\tilde{\zeta}\in \C$ and $z,w\in \C^*$.
The Riemann surface
\be \Sigma\,:\quad H(z,w) = 0\,. \ee
captures all relevant data of the Calabi-Yau geometry.

For example, in the case $X = \CO(-K)\to\F_0$, the mirror curve $\Sigma$ is given by
\be H(z,w) = \sqrt{q_b} \left( z + \frac{1}{z} \right)
+ \sqrt{q_f} \left( w + \frac{1}{w} \right) - 1\,. \label{HF0} \ee
The complex structure moduli $q_b$ and $q_f$ are related to the K\"ahler moduli of $X$ near large volume as $q_{i}\simeq e^{2\pi i t_{i}}$. Upon setting
\be \begin{array}{rl} \sqrt{q_b}&= -\Lambda^2\epsilon\,, \\
  \sqrt{q_f} &= \frac{1}{2}e^{-2\epsilon^2u}\,, \end{array} \label{IIB_SW}
\ee
redefining $w = -1+2\epsilon x$ and $z=\Lambda^{-2}(y+u-x^2)$, and taking $\epsilon\to 0$, the curve $\Sigma$ reduces to the pure $SU(2)$ Seiberg-Witten curve $\Sigma_{SW}^{(N_f=0)}$ from \eqref{Sigma_SW}. The Coulomb branch of the $SU(2)$ theory, parametrized by $u$, is realized as a slice of the complex structure moduli space of $\tilde{X}$, parametrized by $q_b$ and $q_f$.

The compact ``even'' D-branes on the IIA side are all generally mapped to compact D3-branes on $\tilde{X}$. All compact three-cycles $\tilde{\gamma}$ on $\tilde{X}$, however, can be represented as one-cycles%
\footnote{Specifically, any compact three-cycle $\tilde{\gamma}$ on $\tilde{X}$ is an $S^1$ fibration over a disc $D$ in the $(z,w)$-plane with boundary $\gamma=\pd D$ on $\Sigma$. The fibration degenerates on $\gamma\subset D$. These three-cycles have the topology of either $S^2\times S^1$ or $S^3$, and cycles of the two types are (electromagnetically) dual to one another.} %
$\gamma$ on $\Sigma$. Moreover, integrals of the holomorphic three-form $\Omega$ over $\tilde{\gamma}\subset\tilde{X}$ descend to integrals of a (logarithmic) one-form $\lambda$ over the corresponding cycles in $\Sigma$. In the Seiberg-Witten limit, this one-form becomes proportional to $\epsilon$ times the Seiberg-Witten form $\lambda_{SW}$, leading to the usual formula \eqref{Z_gamma} for the central charge of states in $\CN=2$ gauge theory.

Coming back again to the local $\F_0$ example with mirror curve \eqref{HF0}, we expect to see the D0, D2$_{\rm(fiber)}$, D2$_{\rm(base)}$, and D4 branes from the type IIA side represented as one-cycles on $\Sigma$. The surface $\Sigma$ is a torus with four punctures, but the integrals of $\lambda$ around these punctures are not all independent: the punctures come in pairs and yield only two distinct integrals. Therefore, the ``relevant'' homology of $\Sigma$ is $H_1(\Sigma,\Z)\simeq \Z^4$, generated by two punctures alongside the ordinary longitude and meridian ($\alpha$ and $\beta$ cycles) of the torus. This matches the charge lattice on the IIA side, given by the (co)homology of $\F_0$.\footnote{For a more detailed complementary discussion of this correspondence, see \eg\ \cite{ABK-modular}.} In the Seiberg-Witten limit $\epsilon\to 0$, the two pairs of punctures collide and become invisible to $\lambda_{SW}$. The only relevant remaining cycles are linear combinations of the $\alpha$ and $\beta$ cycles on the effectively puncture-less torus $\Sigma_{SW}$. These define the sublattice $\Gamma\subset \Z^4$ of gauge theory charges. (Note, however, that the electric and magnetic gauge theory charges are \emph{not} simply the basic $\alpha$ and $\beta$ cycles, since their intersection number must be 2, while $\ev{\alpha,\beta}=1$; \cf\ a nice discussion of cycles on $\Sigma_{SW}$ in \cite{Lerche-SW}.)

It is well known that the complex structure moduli space of the curve $\Sigma$ contains conifold singularities at points where the discriminant of the equation $H(z,w)=0$ vanishes. For \eqref{HF0}, the discriminant is
\be \Delta = \prod_{\eta_1,\,\eta_2=\pm1}\big(1+2\eta_1\sqrt{q_b}+2\eta_2\sqrt{q_f}\big)\,, \ee
containing four irreducible components. In the Seiberg-Witten limit, only two components survive, yielding
\be \Delta_{SW} = (u+\Lambda^2)(u-\Lambda^2)\,,\ee
which precisely corresponds to the two singular points on the Coulomb branch of $SU(2)$ $N_f=0$ theory. Lifted back up to $\Delta$, these points are the loci where a pure D4 brane (\ie\ a pure magnetic state) and a D4 brane with one unit of D2$_{(\rm fiber)}$ charge (a dyonic state) become massless.

In the preceding discussion, we simply declared that electric and magnetic charges correspond to D2 and D4 branes as in \eqref{Dem}. It is physically clear that the electric charge must correspond to a D2 brane on the fiber $\P^1$, since the latter is what generates the mass of the $W$ boson. It is then also easy to see from an intersection calculation (or the simple fact that four-cycles are dual to two-cycles) that the magnetic charge must correspond at least to a D4 brane. However, it is not clear \emph{a priori} that magnetic charge could not correspond to some bound state of D4, D2$_{\rm(fiber)}$, and D0 branes. (A D2$_{\rm(base)}$ contribution is not allowed, again due to the intersection theory.)

One way to resolve this ambiguity is to analytically continue the quantum-corrected periods on $\tilde{X}$ (computed by use of Picard-Fuchs equations) from the large-volume region around $(q_b,q_f)=(0,0)$ 
to the Seiberg-Witten regime $(q_b,q_f)\simeq(0,\frac14)$.
At large volume, the periods are identified with type IIA D-branes by their leading-order terms:
\begin{align} D0 &\leftrightarrow\;\;\Pi_0 =  1\,, \nno  \\
 D2_{\rm(base)} &\leftrightarrow\;\;\Pi_{b} = \frac{1}{2\pi i}\log(q_b) + O(q_b,q_f)\,, \nno \\
 D2_{\rm(fiber)} &\leftrightarrow\;\;\Pi_{f} = \frac{1}{2\pi i}\log(q_f) + O(q_b,q_f)\,, \nno \\
 D4 & \leftrightarrow\;\;\Pi_4 = \frac{1}{(2\pi i)^2}\log(q_b)\log(q_f) +\frac{1}{6} +O(q_b,q_f)\,. \label{D4charge}
\end{align}
Note that the constant term $\chi(\F_0)/24=1/6$ in \eqref{D4charge} is the natural D0 charge induced on a single D4 brane due to its curvature. Analytically continuing these to the Seiberg-Witten regime, we find
\begin{align}
\Pi_f &= 2C \epsilon\, a(u) + O(\epsilon^3)\,, \nno \\
\Pi_4 &= C\epsilon\, a_D(u) + O(\epsilon^3)\,, \nno
\end{align}
for the same proportionality constant $C$ ($=-\sqrt{2}/i\pi$), confirming the identification \eqref{Dem} of D-branes with gauge theory charges.

In light of the identification \eqref{Dem}, the result of Section \ref{sec:stab} relating the weak coupling spectrum of $SU(2)$ theory to an appropriate subset of the stable vector bundles on $\F_0$ is not too surprising. Indeed, one generally expects that BPS D-branes in a type IIA compactification are described by $\pi$-stable (Bridgeland-stable) objects in the derived category of coherent sheaves on $X$, and these in turn are well-approximated by slope-stable sheaves near large volume. Nevertheless, we saw that the weak-coupling ($|u|\to\infty$) region of the Coulomb branch is \emph{not} quite equivalent to large volume in Calabi-Yau moduli space. It is interesting to note that the two regions do not seem to be separated by any walls of marginal stability that are relevant for gauge theory charges.


\section{Refined wall crossing}
\label{sec:refined}

So far, our approach to wall crossing in gauge theory has been based on examples. We would now like to describe the general formalism underlying the results of Section \ref{sec:hbps}, interpreting identities such as \eqref{puresym}-\eqref{qSW} in terms of the ``quantum wall crossing'' described in \cite{DG}. The proceeding discussion applies equally well to BPS invariants of a Calabi-Yau three-fold as to BPS invariants in gauge theory, realized  via geometric engineering.

\subsection{Refined and quantum}
\label{sec:refq}

We begin by making the precise connection between refined and quantum BPS invariants. In an $\CN=2$ theory, whether gauge theory or effective supergravity in a Calabi-Yau compactification of string theory, the BPS Hilbert space $\CH_{BPS}$ is $(\Gamma\oplus\Z)$-graded by the charge $\gamma\in\Gamma$ and the three-dimensional spin $2J_3\in\Z$ of BPS states. It is therefore convenient to write $\CH_{BPS} = \bigoplus_{\gamma\in\Gamma}\CH_{BPS}(\gamma)$, separating components of different charge. We know of course that this Hilbert space also depends on Coulomb-branch (or vector multiplet) parameters $u$ in a piecewise-constant manner, undergoing discontinuous jumps at walls of marginal stability.

All BPS states in 3+1 dimensions have at least a half-hypermultiplet (\ie\ a hypermultiplet without its CPT conjugate) of spin degrees of freedom corresponding to their center-of-mass position in $\R^3$. Put differently, all BPS multiplets are $\CN=2$ supersymmetric, so they are at least the size of a hyper. Factoring out these center-of-mass degrees of freedom to obtain a reduced Hilbert space $\CH_{BPS}'$,
\be \CH_{BPS} =: \Big(\Big[\frac12\Big]+2[0]\Big)\otimes \CH_{BPS}'\,,\ee
one can calculate a refined index of states as
\be \Omega^\rf(\gamma;u;y) := \Tr_{\CH_{BPS}(\gamma;u)'}(-y)^{2J_3'}=\sum_{n\in\Z}\Omega_n^\rf(\gamma;u)(-y)^n\,. \ee
(The integers $\Omega_n^\rf$ defined in this way are all positive, and typically all but finitely many of them vanish.)
Physically, \emph{refined} simply means keeping track of the spin content, as is done here with the formal variable $y$.
For example, the contributions of a (half) hypermultiplet and a (half) vector multiplet to the refined index are
\be
\begin{array}{r@{\qquad}c@{\qquad}c@{\qquad}c}
 \text{hyper}: & \Omega^\rf = 1 &\text{or}& \Omega_0^\rf = 1\,,\vspace{.15cm} \\
 \text{vector}: & \Omega^\rf = -y-y^{-1} &\text{or}& \Omega_{-1}^\rf = \Omega_{1}^\rf = 1\,.
\end{array} \label{hypvec}
\ee

In order to express wall crossing for the refined indices $\Omega^\rf$ in terms of quantum operators,
one must in general construct an operator $\e{\gamma}$ for every charge $\gamma\in\Gamma$ and a more complicated combination
\begin{align} \bU_\gamma &= \prod_{n\in \Z}\bE\big((-q^{\frac12})^n\e{\gamma})^{(-1)^n\Omega_n^\rf} \label{genU}  \\
 &= 1+ \frac{\Omega^\rf\big(\gamma;-q^{\frac12}\big)}{q^{\frac12}-q^{-\frac12}} \,\e{\gamma}+\ldots \notag
\end{align}
for every \emph{state} of charge $\gamma$ in the BPS Hilbert space.
Here, $\bE (x)$ is the quantum dilogarithm function described in the introduction,
\be \bE(x) := \sum_{n=0}^\infty \frac{(-q^{\frac12}x)^n}{(1-q)\cdots(1-q^n)} = \prod_{i=0}^\infty\big(1+q^{i+\frac12}x\big)^{-1}\,, \ee
and the quantum operators $\e{\gamma}$, $\gamma \in \Gamma$, satisfy relations
\be \e{\gamma}\e{\gamma'} = q^{\frac{\ev{\gamma,\gamma'}}{2}}\e{\gamma+\gamma'}=q^{\ev{\gamma,\gamma'}}\e{\gamma'}\e{\gamma}\,, \label{qtorus} \ee
where $\ev{\gamma,\gamma'}$ is the antisymmetric electromagnetic product of charges.
The algebra generated by the $\e{\gamma}$ is called the {\it quantum torus} and plays a key role in our discussion here as well as in the intrinsically motivic analysis later in Section \ref{sec:math}.

Now, suppose that one defines a local splitting of $\Gamma$ into particles and antiparticles close to a wall of marginal stability. This divides the central charge plane in half, distinguishing a sector of total angle $\pi$ where the central charges of putative particles (versus antiparticles) must lie. Given Coulomb branch (or vector multiplet moduli space) parameters $u_\pm$ on either side of the wall, one can then form composite operators
\be \bA(u_\pm) \,\, = \prod_{{\rm states}\,\gamma \,\in\, \CH_{BPS}(u_\pm)}^\curvearrowright \bU_\gamma\,, \label{genA} \ee
taking a product over all BPS particles (not antiparticles) in order of increasing phase of the central charge. The quantum wall-crossing formula (\cf\ \cite{DG}) states that
\be \boxed{\bA(u_+) = \bA(u_-)}\,. \label{qWC} \ee

Let's illustrate this in the special case of $\CN=2$ super-Yang-Mills theories with gauge group $SU(2)$. The charge lattice $\Gamma$ is two-dimensional, generated by electric and magnetic charges $\gamma_e,\gamma_m$. Therefore, the quantum torus is generated by $\e{e}$ and $\e{m}$, where
\be \e{e}\e{m} = q \e{m}\e{e}\,, \ee
and for a general state of charge $\gamma = n\gamma_e+m\gamma_m$ we have
\be \e{\gamma} = q^{-\frac{nm}{2}}\e{\gamma_e}^{\;n}\e{\gamma_m}^{\;m}\,. \ee
The only spin multiplets that arise in $SU(2)$, $\CN=2$ super-Yang-Mills theories are hypers and vectors.
For every hypermultiplet BPS state of charge $\gamma=n\gamma_e+m\gamma_m$, it follows from \eqref{hypvec} and \eqref{genU} that the wall-crossing operator is
\be \bU_{n,m} := \bU_\gamma = \bE(\e{\gamma})\,, \label{Uhyp} \ee
and for the vector multiplet of electric charge $\gamma = 2\gamma_e$ we find the operator
\be \bU^{\rm vect}_{2,0} = \bE\big(-q^{\frac12}\e{e}^2\big)^{-1}\bE\big(-q^{-\frac12}\e{e}^2\big)^{-1}\,. \ee
The quantum identities \eqref{puresym}-\eqref{qSW} then become direct specializations of the wall-crossing formula \eqref{qWC} for theories with $N_f=0,1,2,3$.

\subsection{Classical limit and factorization}
\label{sec:fact}

We conclude this short section with some brief comments about the factorization of refined/motivic wall crossing factors in physics and mathematics --- leading in to the mathematical analysis of wall crossing in $\CN=2$ gauge theory of Section \ref{sec:math}.

The main conjecture of \cite{DG} is that, for BPS invariants, refined = quantum = motivic. In \cite{KS1} (see also \cite{KS-summary}), motivic operators $A^\mot(u)$, analogous to the $\bA(u)$ appearing here, are defined using methods of motivic integration. In particular, such operators $A^\mot_V(u)$ are defined for every convex sector $V$ (of angle smaller than $\pi$) in the central charge plane, and it is shown that if $V$ is split into disjoint subsectors $V=\bigcup_iV_i$ then
\be A^\mot_V = \prod^{\curvearrowright}_i A^\mot_{V_i}\,, \label{motfact} \ee
where the product is again taken in order of increasing argument of the central charge. In the limit that each $V_i$ contains a single BPS ray, this looks almost like the formula \eqref{genA}.

The complete factorization of motivic invariants into \emph{states} as in \eqref{genU}, \eqref{genA} is suggested by simple examples in \cite{KS1}, and especially by a factorization of the corresponding \emph{classical} invariants obtained in the limit $q^{\frac12}\to -1$. Recall that the operators $\bA$ or $A^\mot_V$ generate automorphisms $T$ of the quantum torus via conjugation:
\be T_V:\;\; \e{\gamma} \mapsto A_V^\mot \e{\gamma}({A_V^\mot})^{-1}\,. \ee
While the $A_V^\mot$ can have poles at $q^{\frac12}\to-1$, it is conjectured in \cite{KS1} that the $T_V$ are regular; in the classical limit $q^{\frac12}\to-1$, they become symplectomorphisms $K_V$ of a classical, symplectic torus. It is shown in \cite{KS1} that these $K_V$'s factorize completely.

In the upcoming work \cite{KS2}, it is also shown that motivic Donaldson-Thomas invariants $A^\mot_V$ have a structure exactly like that of \eqref{genU}, \eqref{genA} in situations where the underlying Calabi-Yau category can be described via representations of a quiver with potential. 

Physically, the decomposition \eqref{genU}, \eqref{genA} is quite natural. The operators $\bA(u)$ can in fact be thought of as encoding the Fock space structure of BPS states on either side of a wall of marginal stability --- this can be seen, for example, by deriving the refined semi-primitive wall-crossing formula of \cite{DG, DenefM} from \eqref{qWC}.
A very interesting interpretation of our full wall-crossing formula \eqref{qWC} in $\CN=2$ gauge theory, based on a chain of string dualities, was also recently proposed in \cite{CV}.

As a final remark, we observe that in order to compare physical refined formulas to motivic invariants one must always choose a finite, positive basis for the set of BPS particles, given a certain particle-antiparticle split. In other words, one needs a basis $\gamma_i$ for $\Gamma^+$, the half of the charge lattice containing particles, such that all BPS states have charges $\gamma = n_i\gamma_i$ with $n_i>0$. It is believed that this is always possible.\footnote{In the mathematical context, the particle-antiparticle split defines a $t$-structure, with the positive basis as its heart.} For example, the wall-crossing formula \eqref{puresym} for pure $SU(2)$ Yang-Mills theory in a positive basis $\gamma_1=2\gamma_e-\gamma_m, \gamma_2=\gamma_m$ (now with $\ev{\gamma_1,\gamma_2}=2$) looks like
\be N_f=0:\qquad \bU_{1,0}\bU_{0,1} = \bU_{0,1}\bU_{1,2}\bU_{2,3}\cdots\bU^{\rm vect}_{1,1}\cdots\bU_{3,2}\bU_{2,1}\bU_{1,0}\,. \label{SW0pos} \ee
Given a particle-antiparticle split and an ordering of the central charge, the existence of a finite, positive basis guarantees unique factorization of a composite operator $\bA(u)$ into operators $\bU_\gamma$. Further implications of this for the structure of physical moduli spaces will be discussed in \cite{DJ}.


\section{Motivic wall crossing}
\label{sec:math}

In the previous sections we discussed wall crossing for refined BPS invariants, and began to look at how refined/quantum wall crossing is related to motivic wall crossing.
It was proposed in \cite{DG} that refined BPS invariants in physics
are equal to the motivic BPS invariants of \cite{KS1}, where the spin content
of $\Omega^{{\rm ref}}$ is identified with the ``motivic'' content of $\Omega^{{\rm mot}}$,
\be \Omega^\rf(\gamma;y) = \Omega^{\rm mot}(\gamma;q)\,,\qquad y \leftrightarrow -q^{\frac12}\,. \ee
In this section, we present further evidence for this identification by analyzing
motivic BPS invariants in the simple examples related to $\CN=2$ supersymmetric gauge theories
with $SU(2)$ gauge group and $N_f < 4$ matter hypermultiplets in the fundamental representation.
In particular, in the case of pure $\CN=2$ super-Yang-Mills theory we find that
motivic BPS invariants obey the same wall-crossing formula \eqref{puresym}
as the refined BPS invariants, and we comment on the examples related to $\CN=2$ super-QCD.

In the case of these $\CN=2$ gauge theories, there are actually several ways to derive the motivic wall-crossing formulas. We first present an approach based on representations of quivers that directly invokes the methods of \cite{KS1}, and then consider a complementary approach based on the so-called ``cohomological Hall algebra,'' which will appear in the upcoming work \cite{KS2}. The latter strongly indicates that motivic Donaldson-Thomas invariants indeed satisfy the factorization properties discussed in Section \ref{sec:fact}.

\subsection{Wall crossing and quivers}

Recall \cite{KS1, KS-summary} that motivic Donaldson-Thomas invariants are formulated in the context of ind-constructible 3d Calabi-Yau categories. Physically, such a category $\CC$ should be thought of as the category of D-branes on a given Calabi-Yau three-fold. The $K$-theory $K_0(\CC)$ is mapped to the charge lattice $\Gamma$, and these categories are endowed with an additive map $\CZ:\Gamma\to \C$ corresponding to the central charge, which is part of the mathematical definition of the stability condition that singles out BPS objects.

To a 3d Calabi-Yau category $\CC$, one associates a quantum torus $\CR(\CC)$ over the coefficient ring, which is the localization of the ring of motivic stack functions on the space of objects (D-branes) in $\CC$. It specializes precisely to the quantum torus described in Section \ref{sec:refq}, with relations \eqref{qtorus}. Given a convex sector $V$ in the central charge plane, motivic Donaldson-Thomas invariants $A_V^\mot$ are defined based on the theory of motivic integration (and in particular based on the notion of motivic Milnor fiber of the ``superpotential'' associated to objects in $\CC$). The abstract quantum variable $q$ corresponds to the motive of the affine line $\L$.  As we have seen (in the case of the quantum operators $\bA(u)$), these invariants are invertible elements of a suitable completion of the quantum torus $\CR(\CC)$.

\begin{figure}[htb]
\centering
\includegraphics[width=2in]{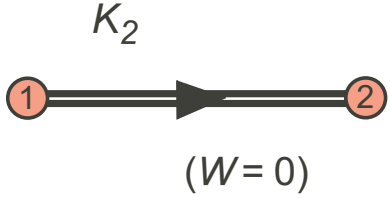}
\caption{The quiver $K_2$ for $SU(2)$ super-Yang-Mills with $N_f=0$.}
\label{fig:K2}
\end{figure}

In the case of pure $SU(2)$ super-Yang-Mills theory, we saw that states in the gauge theory are identified with a subset of the D-branes on the noncompact Calabi-Yau $X = \CO(-K)\to \F_0$. For this subset, the corresponding category $\CC$ can also be thought of as a category of representations%
\footnote{Physically, the relation between $SU(N)$ gauge theories and quivers is explained \eg\ in \cite{Fiol-parton} (see also \cite{DFR-noncpt}). It is interesting to note that the ``natural'' stable representations of quivers, corresponding to a small-volume limit of Calabi-Yau manifolds, make manifest the finite LHS of the wall-crossing formulas --- in contrast to the large-volume analysis of stable bundles (Section \ref{sec:stab}), which produces the infinite RHS.} %
of the Kronecker quiver $K_2$ (Figure \ref{fig:K2}). The motivic wall-crossing formula can then be derived from the representation theory of this quiver, which we briefly review.%

The Kronecker quiver $K_2$ has two vertices $1,2$ and two arrows from $1$ to $2$. Representations consist of vector spaces $V_i$ assigned to each node and maps $\phi_j:V_1\to V_2$ assigned to each arrow. The dimension vector $\gamma = (\dim V_1,\dim V_2)$ of a representation corresponds to the electric and magnetic charges of states written in the positive basis $(\gamma_1,\gamma_2)$, as described at the end of Section \ref{sec:fact}. (Thus $\gamma=(a,b)$ corresponds to $2a\gamma_e+(b-a)\gamma_m$.) We will be looking for stability conditions on the derived category of finite-dimensional representations of $K_2$. More precisely, we fix the $t$-structure given by finite-dimensional representations of $K_2$ and vary the central charge $\CZ$. The central charge of a quiver representation with dimension $\gamma=(a,b)$ can be identified with a pair of non-negative integers $(\Theta_1,\Theta_2)$, in terms of which the representation has slope $\mu = \frac{a\,\Theta_1+b\,\Theta_2}{a+b}$. The notion of stability can be defined via this slope function in the standard way: a representation $R$ is (semi)stable if for all nontrivial proper subrepresentations $R'$ one has
\be \mu(R') \underset{\text{\bf\tiny(---)}}{<} \mu(R)\,. \ee

It is easy to see that besides the trivial stability condition, there are only two other distinct choices: $\Theta=(0,1)$ and $\Theta=(1,0)$. In the case of the former there are only two stable objects, which are irreducible representations of dimensions $(0,1)$ and $(1,0)$ (\ie\ the monopole and dyon stable at strong coupling), and any semistable is the direct sum of copies of either of these.

In the case of the stability condition $(1,0)$ the picture is more complicated, but the classification of stable and semistable representations is well known (see \eg\ \cite{Re1} Sec. 6.1 or \cite{DFR-noncpt} App. A). First, one observes that any semistable representation is a direct sum of indecomposable representations of $K_2$ of the same slope. Second, one uses the classification of indecomposable representations (which is basically due to Kronecker) in order to obtain the description of semistable representations. Here is the result:
\begin{itemize}

\item[a)] There is a unique stable representation for any dimension vector $(k,k+1)$ or $(k+1,k)$. In the equivalent geometric description these correspond to morphisms between ample sheaves $O(k)$ and $O(k+1)$ on $\P^1_b$ (which are pulled back to $\F_0=\P^1_b\times P^1_f$ to describe dyonic D4 branes). 
Semistable representations that have the corresponding slopes $\frac{k}{2k+1}, \frac{k+1}{2k+1}$ are direct sums of stable ones (they are also so-called polystable), since the dimensions are relatively prime to the stability condition. Each stable representation $E_k$ of the one of these dimensions is a Schur object in the category of representations, \ie\ its automorphism group is $\C^{\ast}$, and there are no $\Ext^i(E_k,E_k)$ for $i<0$. Furthermore $\Ext^1(E_k,E_k)=0$. Thus, for a semistable object $nE_k=E_k^n$ of one of these slopes the only non-zero group $\Ext^i(nE_k,nE_k)$ with $i\le 1$ appears for $i=0$, and it is isomorphic to ${\rm Mat}(n,\C)$.

\item[b)] We have a $\P^1$-family of stable representations of dimension vector $(1,1)$, and no stable representations of dimension vector $(k,k)$ for $k\geq 2$. For $\gamma=(1,1)$ every semistable representation is stable. In contrast to case (a), the stability condition is not coprime to $\gamma=(k,k)$, and
there is a $\P^1$-family of indecomposable semistable representations for each dimension $(k,k)$. One can think that the first arrow in the $K_2$ quiver is represented by the identity map, while the second one is represented by the $k\times k$ Jordan block with eigenvalue $\lambda\in \C^{\ast}$ (one can upgrade $\lambda$ to a point on $\P^1$). In the geometric description such an object corresponds to the torsion sheaf with the support at a point $\lambda\in \P^1_b$ (which can be pulled back to become the electric D2 brane supported on the fiber $\P^1_f$ of $\F_0$).
Then the moduli space of semistables of dimension $(k,k)$ can be identified with the moduli space of torsion sheaves on $\P^1$ of length $k$.

\end{itemize}

The factorization formula \eqref{motfact} expresses the motivic Donaldson-Thomas invariant $A_V^\mot$ as a product of invariants $A_{\ell(\mu)}^\mot$ associated to BPS rays $\ell(\mu)$ with slopes $\mu\in V$. The above considerations (a) and (b) imply that with stability condition $\Theta=(0,1)$ the only nontrivial factors $A_{\ell(\mu)}$ entering this product occur for slopes $\mu=\frac12,\,\mu = \frac{k}{2k+1},$ and $\mu = \frac{k+1}{2k+1}$, $k=0,1,2,...$.

For representations described in (a), the abelian category ${\cal C}_\ell$ corresponding to the ray $\ell$ is generated by one Schur object $E_k$, of dimension either $(k,k+1)$ or $(k+1,k)$ depending on the slope. The computation of $A_\ell^\mot$ in these cases is completely analogous to the one for the category generated by one spherical object from \cite{KS1}, Section 6.4.
More precisely, recall that the explicit formula from \cite{KS1} can be schematically written as
\be A_\ell^\mot=\sum_{[E]}{w(E)\over {[{\rm Aut}(E)]}}\,\e{cl(E)}\,, \ee
where the sum is taken over the isomorphism classes of objects of the category generated by semistables that have slope belonging to the ray $\ell$, the quantum-torus element $\e{cl(E)}$ corresponds to the ($K$-theory) class of $E$, and $w(E)$ is a certain motivic weight. In the case of a quiver the weight is equal to $q^{{\dim\,\Hom(E,E)-\dim\,\Ext^1(E,E)\over{2}}}$. In case (a), however, we have $\Ext^1(E,E)=0$, and the isomorphism class of any object $E=nE_k$ is uniquely determined by its dimension. This means that the motive of representations of the dimension $n(k,k+1)$ (resp. $n(k+1,k)$) is trivial.  Therefore, the motivic weight is equal to $q^{n^2/2},\, n=\dim\,\Hom(nE_k,nE_k)$. Together with the formula $[{\rm Aut}(nE_k)]=[GL(n)]=(q^n-q^{n-1})...(q^n-1)$ (and $\e{cl(nE_k)}=\e{cl(E_k)}^n$) this gives $A_\ell^\mot$ as the quantum dilogarithm function $\bE\big(\e{cl(E_k)}\big)$.

In the case (b) all semistables have slope $1/2$. Recall that they correspond to torsion sheaves on $\P^1$. The Euler pairing between $K$-theory classes of two torsion sheaves is trivial, so $w(E)=1$. On the other hand, the isomorphism class of the torsion sheaf is no longer determined by its dimension, so we will have a non-trivial motive of such sheaves with a fixed class in the $K$-theory (or, equivalently, with fixed dimension of the representation).
An important fact, however, is that $A_\ell^\mot$ factorizes as the product of the  generating functions for torsion sheaves on $\C$ and torsion sheaves concentrated at one point, \eg\ at zero. Indeed, torsion sheaves concentrated at different points are orthogonal in the derived category with respect to the $\Ext^{\bullet}$ pairing.
Thus we can write $A_{\ell(1/2)}^\mot=A_\C^\mot A_{0}^\mot$, with factors corresponding to torsion sheaves on $\C$ and torsion sheaves supported at $0\in \C$.

The torsion sheaf on $\C$ of length $n$ is the same as the $n$-dimensional representation of the algebra $\C[x]$. Isomorphism classes are parametrized by $n\times n$ matrices, hence
the  motive of sheaves of length $n$  is $q^{n^2}$ (equivalently one can count ${\bf F}_q$-points of the scheme of finite-dimensional representations of the algebra ${\bf F}_q[x]$, where ${\bf F}_q$ is the finite field with $q$ elements). Thus we see that
\be A_\C^\mot=\sum_{n\ge 0} \frac{q^{n^2}}
{(q^n-q^{n-1})\dots(q^n-1)} t^n\,, \ee
where $t=\e{cl(E_1)}=\e{(1,1)}$ is the class of the representation $(1,1)$.
It is easy to rewrite this series as
\be A_\C^\mot = \sum_{n\ge 0} \frac{q^{n^2/2}}
{(1-q)\dots(1-q^n)} (-q^{1/2}t)^n
 = \bE\big(-q^{1/2}t\big)^{-1}\,.
\ee

In order to compute $A_{0}^\mot$ one observes (by the same reasoning as before) that
$A_\C^\mot=A_{0}^\mot A_{\C^{\ast}}^\mot$, where the second factor corresponds to the moduli space of torsion sheaves on $\C^{\ast}$. In order to compute the latter, one uses the fact that a torsion sheaf on $\C^{\ast}$ of length $n$ corresponds to an $n$-dimensional representation of the algebra $\C[x]$ for which the variable $x$ is represented by an invertible matrix. Such representations form the motive $[GL(n)]$, hence
\be A_{\C^{\ast}}^\mot=\sum_{n\ge 0}\frac{[GL(n)]}{[GL(n)]}t^n=\sum_{n\ge 0}t^n=\frac{1}{1-t}\,. \ee
Therefore
\be A_{0}^\mot=A_\C^\mot(1-t)=\prod_{i\ge 0}{(1-q^{i}t)}=\bE\big(-q^{-1/2}t\big)^{-1}\,,
\ee
and we finally conclude that
\be A_{\ell(1/2)}^\mot=\bE\big(-q^{1/2}t\big)^{-1}\bE\big(-q^{-1/2}t\big)^{-1} = ``\bU^{\rm vect}(t)\text{''}\,. \ee

Altogether, taking the sector $V$ to have angle sufficiently close to $\pi$ in the central charge plane (\ie\ including all the slopes we have just discussed), and equating the factorization of $A_V^\mot$ into BPS rays for the stability parameter $\Theta =(1,0)$ to the factorization for $\Theta = (0,1)$, we have derived from first principles the quantum/motivic wall-crossing formula for pure $SU(2)$ theory in the form \eqref{SW0pos}

\begin{rmk} Another way to derive the same quantum wall-crossing formula is to use the approach of \cite{KS1}, Section 7 and \cite{Re1}, where the difference equation $A_\ell(qt)=F_\ell(t)A_\ell(t)$ was studied. The idea is to interpret $F_\ell(t)$ as the generating function of framed cyclic representations of the quiver $K_2$ (see \cite{Re1} for the definitions).
\end{rmk}

\begin{figure}[htb]
\centering
\includegraphics[width=2in]{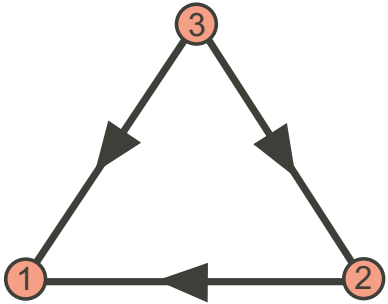}
\caption{The acyclic quiver for $N_f=1$ $SU(2)$ theory.}
\label{fig:N1}
\end{figure}

The above considerations as well as the techniques of \cite{Re1} (and \cite{KS1}, Section 7) can be generalized to other quivers which appear in $\CN=2$ gauge theory. For example, in the case of $SU(2)$ super-QCD with $N_f=1$, the geometry of the surface $\F_0$ blown up at one generic point leads to a quiver with three vertices and three (single) arrows, as shown in Figure \ref{fig:N1}. The potential is trivial. The charges $\gamma_1,\gamma_2,\gamma_3$ associated to the three vertices are related to gauge theory charges as $\gamma_1=\gamma_m$, $\gamma_2=\gamma_e$, and $\gamma_3=\gamma_e-\gamma_m$.

There is an obvious stability condition, for which the motivic DT-invariant is given by the product of three quantum dilogarithm functions corresponding to the three vertices of the quiver (this follows e.g. from \cite{KS1}, Proposition 16). Of course, these are the three stable BPS states in the strong-coupling region of $N_f=1$ theory.

Now, a stable representation of a given slope $\mu$  is indecomposable, and since our quiver is of extended Dynkin type, indecomposable representations correspond to positive roots of the quiver. In order to reproduce the RHS of the $N_f=1$ wall-crossing formula, we would like to find a stability condition for which all such indecomposable representations appear in the product. Markus Reineke\footnote{We thank to him for kindly sending us the results of his computations.} has suggested to use the central charge $\CZ(\gamma_1,\gamma_2,\gamma_3)=-(\gamma_2+\gamma_3)+i(\gamma_1+2\gamma_2+\gamma_3)$, which means that the slope function is
$\mu(\gamma_1,\gamma_2,\gamma_3)=(\gamma_2+\gamma_3)/(\gamma_1+2\gamma_2+\gamma_3)$.
The moduli spaces of stable representations are
either empty, or single points, or an affine line (for dimension vector
$(1,1,1)$), similar to the case of $K_2$.
In the classical limit $q^{\frac12}\to-1$, the product on the RHS of the wall-crossing formula was obtained by Reineke via \cite{Re1}. He defines automorphisms of $\Q[[x_1,x_2,x_3]]$ as:
\be K_{\gamma}(x_i) = x_i(1+\sigma x^{\gamma})^{\ev{\gamma_i,\gamma}}\,, \ee
where $\gamma=(a,b,c)=a\gamma_1+b\gamma_2+c\gamma_3$, and the number $\sigma$ is $-1$ if $\gamma_1+\gamma_2+\gamma_3$ is divisible by $3$, and $+1$ otherwise.
He then arrives at a classical product formula in the form
\be
K_{001} K_{010} K_{100} = K_{100} K_{110} K_{211} K_{221} K_{322} K_{332} \cdots \CK \cdots K_{233} K_{223} K_{122} K_{112} K_{011} K_{001}\,,
\ee
where $\CK = (K_{111})^{-2} K_{101} K_{010}$.
This is indeed (a slight generalization of) the classical wall-crossing formula for $N_f=1$ theory (\cf\ \cite{GMN-II}), as can be checked by converting to charges $\gamma_e$ and $\gamma_m$. It can be ``upgraded'' to the quantum case in a fairly straightforward manner via the same methods we presented above for $N_f=0$ theory.

\begin{rmk} In a very interesting recent paper \cite{GPS}, the ``classical''
product formula was interpreted in a completely different fashion, by using relative Gromov-Witten theory.
The exponents (``classical DT-invariants'') appear there as the number of rational curves in $\P^2$
which pass through given points on two given divisors and have a prescribed tangency order at the third divisor
``at infinity''.
It would be interesting to understand the quantum analog of this result.
\end{rmk}

\subsection{Factorization and cohomological Hall algebra}

In this final section, we briefly discuss another approach to motivic DT-invariants suggested in \cite{KS2}.
The results of  \cite{KS2} are proved for arbitrary quivers with potential. Every such quiver gives rise to a $3d$ Calabi-Yau category. We reproduce here a very special
case of the general theory of \cite{KS2} in the case of a quiver with trivial potential.

In \cite{KS2}, Kontsevich and the third-named author (Y.S.) introduce a new structure of associative unital algebra
(called cohomological Hall algebra) on the $\Z_{\ge 0}^I$-graded abelian group
\be A:=\oplus_\gamma A_\gamma\,\,, \ee
where each component is defined as an equivariant cohomology
\be A_\gamma:=H^\bullet_{G_\gamma} (M_\gamma) \ee
of the vector space $M_{\gamma}$ of quiver representations of fixed dimension $\gamma\in\Z_{\ge 0}^I$, and $G_{\gamma}$ is the complexified quiver ``gauge group'' acting on $M_{\gamma}$. The algebra product (which we do not recall here in full generality) admits a very
explicit form via torus localization with respect to the maximal torus $T_{\gamma}\subset G_{\gamma}$.

As an example, let $Q_d$ be the quiver with one vertex and $d$ loops.
Then $A$ is the sum of equivariant cohomology of the point, with appropriate shifts: $A\simeq H^{\bullet}(BGL(n),\C)[(d-1)n^2]$. Elements of the cohomological Hall algebra can then be identified with polynomials, and the explicit formula for the product reads:
\begin{align}
&(f_1\cdot f_2)(x_1,\dots,x_{n+m}):= \notag\\
&\qquad\sum_{\substack{ i_1<\dots<i_n\\ j_1<\dots<j_m\\
\{i_1,\dots,i_n,j_1,\dots,j_m\}=\\
=\{1,\dots,n+m\}}} f_1(x_{i_1},\dots,x_{i_n})\,f_2(x_{j_1},\dots,x_{j_m})\,
\left(\prod_{k=1}^n\prod_{l=1}^m(x_{j_l}-x_{i_k})\right)^{d-1}
\end{align}
for symmetric polynomials $f_1$ (in $n$  variables) and  $f_2$ (in $m$ variables). The product $f_1\cdot f_2$ is a symmetric polynomial in $n+m$ variables.
The algebra has a double grading such that a homogeneous symmetric polynomial of degree $K$ in $n$ variables
 has bigrading $(n,2K+(1-d)n^2)$ (the bigrading corresponds to the shifts by $(d-1)n^2$ above).
The resulting Hilbert-Poincar\'e series is
$$E_d(z,q^{1/2})=\sum_{n\ge 0, m\in \Z}\dim(A_{n,m})\,z^n q^{m/2}=\sum_{n\ge 0} \frac{q^{(1-d)n^2/2}}
{(1-q)\dots(1-q^n)} z^n\in \Z((q^{1/2}))[[z]]\,\,.$$

The results of \cite{KS2} imply the following factorization formula:
\begin{prp} For any $d\ge 0$ there exist integers $\delta(n,m)=\delta^{(d)}(n,m)\ge 0$ for all $n\ge 1$ and $m\in (d-1)n+2\Z=(1-d)n^2+2\Z$, such that for a given number $n$
 we have $\delta(n,m)\ne 0$ only for finitely many values of $m$, and
\be
E_d(z,q^{1/2})= \prod_{n\ge 1} \prod_{m\in \Z}
\bE\big((-q^{\frac12})^{m-1}z^n\big)^{(-1)^{m-1} \delta(n,m)}
\,.
\ee
\end{prp}

Notice that $E_d(z,q^{-1/2})$ coincides with the motivic DT-invariant of the $3d$ Calabi-Yau category associated with the quiver $Q_d$ (see \cite{KS1}, Section 8 about the details of the correspondence between $3d$ Calabi-Yau categories and quivers with potentials). The above factorization formula is generalized in \cite{KS2} to more general quivers with non-zero potentials. It is shown there that by introducing a stability condition on the category of finite-dimensional representations of $Q$ one obtains a factorization formula for the motivic DT-invariant of the corresponding $3d$ Calabi-Yau category.
These factorizations can be thought of as a further generalization of the motivic wall-crossing formulas, and coincide with the ``expected''  factorization of refined/quantum invariants \eqref{genU}, \eqref{genA}.  They also generalize the results of  \cite{Re1, Re2}.
Applying the results of \cite{KS2} to (\eg) the Kronecker quiver $K_2$ leads to yet another derivation of the quantum/motivic wall-crossing formulas for super-Yang-Mills theory.


\acknowledgments

We would like to thank D. Jafferis, M. Kontsevich, A. Neitzke, and M. Reineke for very useful discussions,
and S. Deser and C. Silberstein for providing important clues.
The work of SG is supported in part by DOE Grant DE-FG03-92-ER40701,
in part by NSF Grant PHY-0757647, and in part by the Alfred P. Sloan Foundation.
Y.S. is grateful to IHES for excellent research conditions. His work was partially supported by an NSF grant.
Opinions and conclusions expressed here are those of the authors
and do not necessarily reflect the views of funding agencies.


\bibliographystyle{JHEP_TD}
\bibliography{SW-Motivic}

\end{document}